\documentclass[10pt,conference]{IEEEtran}
\usepackage{cite}
\usepackage{amsmath,amssymb,amsfonts}
\usepackage{algorithmic}
\usepackage{graphicx}
\usepackage{textcomp}
\usepackage{float}

\usepackage{xcolor}
\def\BibTeX{{\rm B\kern-.05em{\sc i\kern-.025em b}\kern-.08em
    T\kern-.1667em\lower.7ex\hbox{E}\kern-.125emX}}

\usepackage{multirow}
\usepackage[breaklinks=true]{hyperref}
\def\CC{{C\nolinebreak[4]\hspace{-.05em}\raisebox{.2ex}{++}}}

\usepackage{caption}
\usepackage{listings}

\lstdefinestyle{mystyle}{
    captionpos=b,
    basicstyle=\ttfamily\footnotesize,
    frame=single
}

\begin{document}

\title{Taming the Beast: Fully Automated Unit Testing with Coyote \CC}

\author{
\IEEEauthorblockN{Sanghoon Rho}
\IEEEauthorblockA{\textit{CODEMIND Corporation}\\
Seoul, South Korea \\
rho@codemind.co.kr}\\
\IEEEauthorblockN{Yeoneo Kim}
\IEEEauthorblockA{\textit{CODEMIND Corporation}\\
Seoul, South Korea \\
yeoneo@codemind.co.kr}\\
\and
\IEEEauthorblockN{Philipp Martens}
\IEEEauthorblockA{\textit{CODEMIND Corporation}\\
Seoul, South Korea \\
philipp.m@codemind.co.kr}\\
\and
\IEEEauthorblockN{Seungcheol Shin}
\IEEEauthorblockA{\textit{CODEMIND Corporation}\\
Seoul, South Korea \\
shin@codemind.co.kr}\\
}

\maketitle

\begin{abstract}
In this paper, we present Coyote \CC, a fully automated white-box unit testing tool for C and \CC. Whereas existing tools have struggled to realize unit test generation for \CC, Coyote \CC\ is able to produce high coverage results from unit test generation at a testing speed of over 10,000 statements per hour. This impressive feat is made possible by the combination of a powerful concolic execution engine with sophisticated automated test harness generation. Additionally, the GUI of Coyote \CC\ displays detailed code coverage visualizations and provides various configuration features for users seeking to manually optimize their coverage results. Combining potent one-click automated testing with rich support for manual tweaking, Coyote \CC\ is the first automated testing tool that is practical enough to make automated testing of \CC\ code truly viable in industrial applications.
\end{abstract}


\section{Introduction}

Test case generation has been researched for a long time to automate white box dynamic testing, which involves meticulously examining and validating source code of a given software. If this technology becomes a reality, it will not only significantly reduce the cost and effort of software testing by automating the most rigorous testing, but also greatly contribute to ensuring software reliability.

The success of such research can be evaluated by how much code coverage is achieved through automatically generated test cases. Various techniques for test case generation have been studied, including symbolic-execution-based approaches~\cite{klee,cute}, search-based approaches~\cite{EvoSuite}, and LLM-based approaches~\cite{Transformer,schafer2023empirical}, which all compete for higher coverage and performance and each have distinct advantages and disadvantages. There has been a lot of promising research into unit test generation for Java and C and even a tool competition for unit test generation of Java. For \CC\ however, due to its well-known complexity, there has been comparatively little research into unit test generation and the existing research has not been able to report big successes yet.

This paper presents Coyote \CC, a concolic-execution-based fully automated unit testing tool for C/\CC\ that can detect various runtime errors by automatic injection of assertions and achieves high coverage results even for complex \CC\ projects. Coyote \CC\ is an industrial strength tool that is already actively being used in validation of real-world projects such as automotive software. The tool also includes additional features that enable requirement-based test case generation and allow users to manually improve coverage, but the focus of this paper is on its one-click automated test case generation capabilities.

The remainder of this paper is structured as follows. First, we briefly refer to some previous research on unit test generation tools for \CC. We then describe the overall implementation of Coyote \CC\ and challenging problems that were encountered when dealing with \CC. After demonstrating how Coyote \CC\ can be used in practice we also present test results comparing to the latest research for \CC\ projects. Finally we conclude by discussing future works.

\section{Related Works}

While a number of automated white-box unit testing tools exist for Java~\cite{EvoSuite,jcute}, C~\cite{Maist,cute,dart}, and some other languages~\cite{Pex,cuter,jalangi}, there has not been much success yet when targeting \CC, as the language's many intricacies like templates or namespaces greatly complicate automated testing. To make matters worse, out of the already few tools that do claim to be able to handle \CC\ code, some are not available for public use and others ultimately support only a small subset of \CC\ and can thus only deal with very simple programs.

Recently, the two most promising tools for automated unit testing of \CC\ programs have been CITRUS~\cite{Citrus} and UTBot\cite{Utbot}. CITRUS combines concolic execution with fuzzing and employs mutation techniques for test case generation. Typically, CITRUS is executed with a fixed time budget per project for generating parametric test cases and then a fixed time budget for applying libfuzzer to each test case, with the whole testing process reported to take about 10-20 hours to achieve good coverage for projects between 1000 and 20,000 line of code. This large time consumption makes it hard to use CITRUS in many practical applications and especially in the context of continuous testing. UTBot also uses concolic execution for automated test case generation and has good support for C programs since it was developed using the well-established KLEE~\cite{klee} symbolic execution engine as its foundation. However, it only supports a very limited subset of the \CC\ syntax, making it unsuitable for testing most real-world projects.

\section[The Internals of Coyote C++]{The Internals of Coyote \CC}

Coyote \CC\ realizes automatic testing through automatic harness generation and concolic execution to generate test input values. For a more detailed overview about the testing process as a whole, please refer to our previous paper~\cite{Coyote}.

\subsection{Implementation of Concolic Execution}

Coyote \CC\ is based on our own concolic execution engine for automated test generation of C and \CC\ programs. To achieve high coverage results in a relatively short time for programs with a wide range of complex language features, a number of influential design decisions were made during its implementation~\cite{Coyote}, the most significant of which will be summarized in the following.

First, we employ offline symbolic execution, which means that symbolic execution is performed after each individual concrete execution finished, in contrast to online symbolic execution which runs at the same time as concrete execution in an interleaved fashion. Offline mode tends to have reduced memory usage compared to online mode and is thus especially advantageous when symbolic execution is run in a highly parallel implementation of test case generation. Second, the insertion of tracing code and the symbolic execution itself are not performed on the source code level but rather on LLVM IR level, allowing for more detailed tracing and thus more accurate symbolic execution. Third, in order to reach high coverage results in a short amount of time, we use a variant of CCS (Code Coverage Search)\cite{SymExecSurvey} as our default search strategy. A second depth first search (DFS) based strategy is used in cases where the coverage achieved by CCS is not yet sufficient. Last, our approach to symbolic memory modelling uses concrete addresses when writing and symbolic addresses when reading, motivated by the need to find a good balance between solving efficiency and the accuracy of path constraints produced by symbolic execution.

\subsection[Taming the Complexity of C++]{Taming the Complexity of \CC}

To handle the complexity of \CC\ in concolic-execution-based testing, one needs not only a powerful symbolic execution engine but also sophisticated techniques for automated test harness generation. In the following, we discuss common challenges in automated test harness generation for \CC, and give some insight into how these are dealt with by Coyote \CC.

When compared to C, the addition of classes in \CC\ greatly complicates the generation of driver and stub functions. Not only do fields have to be initialized and member functions tested, but many rules regarding the implicit generation or deletion of default/copy/move constructors also make it difficult to simply create and pass around class instances in the test harness. In order to correctly handle classes, it is important to analyze the structure and connection of classes in the source code. In Coyote \CC, this extraction of information about the source code is performed by a custom-built clang plugin in a preliminary step before the actual harness generation. This data is then further processed to obtain more refined information about the program such as call- or use-graphs.

Concolic execution requires the injection of tracing code into the original program to gather information about its concrete execution. However, on the source code level, inserting tracing code into \CC\ programs can be challenging due to unintuitive control flow caused by language features like operator overloading and copy/move constructors, as well as complicated typing in the presence of templates. Coyote \CC\ circumvents these problems by inserting tracing code not into the source code itself but rather into the LLVM IR code generated from it. This greatly reduces the difficulty of injecting tracing code and makes it trivial to guarantee that the inserted tracing code does not cause any undesired side effects.

\begin{figure}
\noindent
\begin{minipage}{\linewidth}
\begin{lstlisting}[language=C++, caption=A sample driver generated by Coyote C++, label={lst:sampledriver}]
Point Point::bound(Point min, Point max) {...}

void __COYOTE_DRIVER_Point_bound() {
  ...
  ::Point x1(__CYC__);
  ::Point x2(__CYC__);
  ::Point cls(__CYC__);
  __COYOTE_SYM_Point(2, (void*)(::Point*)&(cls));
  __COYOTE_SYM_Point(4, (void*)(::Point*)&(x1));
  __COYOTE_SYM_Point(6, (void*)(::Point*)&(x2));
    
  cls.bound(x1, x2);
  ...
}

void __COYOTE_SYM_Point(int id, void *_x) {  
  ::Point *x = (::Point*)_x;
  __COYOTE_ID_SYM_I32(id, (int*)&(x->x));
  __COYOTE_ID_SYM_I32(id+1, (int*)&(x->y));
  ...
}
\end{lstlisting}
\end{minipage}
\end{figure}

\subsection{Test Harness Generation in Practice}

Test harness generation comprises the generation of code for driver functions which initialize input values before calling the functions under test, as well as stub functions that simulate unit-external functions in a way suitable for testing. In the following, we will illustrate how Coyote \CC\ automatically generates test harness code by discussing the test driver generated for a sample function (see listing \ref{lst:sampledriver}). The function to be tested here, \verb|Point::bound|, is a member function of a simple class called \verb|Point| with two integer fields \verb|x| and \verb|y|. As the name suggests, the function receives a maximum and a minimum \verb|Point| as arguments and adjusts the values of the \verb|this| object's \verb|x| and \verb|y| fields so that they lie between those of the argument points.

In order to automatically test this member function, the test driver function has to initialize the arguments \verb|min| and \verb|max| as well as the 
\verb|this| object that the member function is invoked on before actually calling the member function. To this end, next to the main driver function \verb|__COYOTE_DRIVER_Point_bound|, Coyote \CC\ also generates a helper function \verb|__COYOTE_SYM_Point| that handles the symbolic initialization of \verb|Point| instances. The individual fields \verb|x| and \verb|y| are then initialized through calls to \verb|__COYOTE_ID_SYM_I32|, which is a function provided by the Coyote \CC\ API that initializes its second argument to a 32bit integer value associated to the symbol ID given as its first argument. While the \verb|Point| class in this sample code only has two fields, classes in real-world \CC\ programs often have dozens of fields that may each in turn be of class types containing more fields recursively. If the initialization code for such classes is not carefully split into separate functions, driver code alone can easily grow to take up tens or hundreds of megabytes.

Note also that the \verb|Point| instances are initially created through a stub constructor that was added by Coyote \CC\ as a replacement for the class's default constructor, which got implicitly deleted because \verb|Point| has a user defined constructor. As . We avoid the invocation of such user defined constructors in test harness code because the test harness code should focus only on the current function under test\footnote{As user defined constructors are also test targets, they are of course still called from their respective driver functions.}.

\section[Using Coyote C++]{Using Coyote \CC}

In this section, we give an overview of how Coyote \CC\ can be used in practice for fully automated unit testing of \CC\ projects. For more details and concrete step-by-step guidance for the usage of Coyote \CC\ please refer to the user manual provided alongside the demo version of Coyote \CC.

\subsection{Executing Tests}

\begin{figure}
    \centering
    \includegraphics[width=\linewidth]{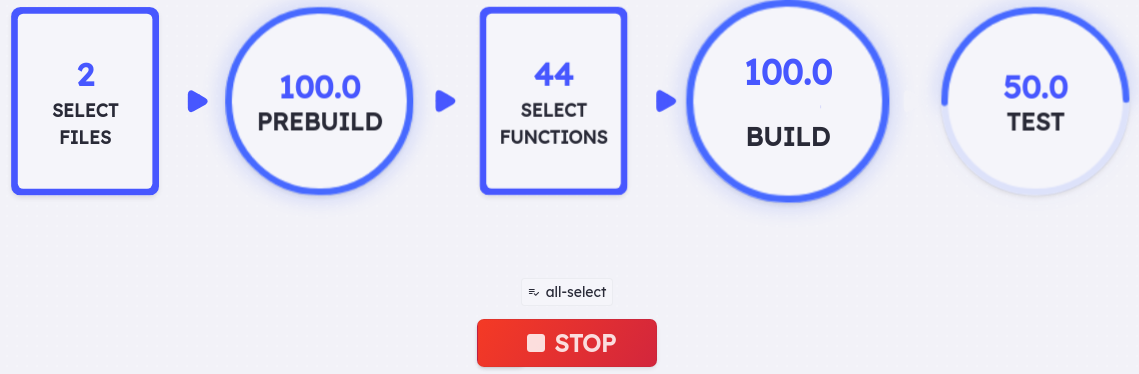}
    \caption{Progress displayed in the main page of Coyote C++}
    \label{fig:mainscreen}
\end{figure}

After finishing the initial setup for a project, involving e.g. the specification of include paths, simply clicking the "Run" button in the main page of Coyote \CC\ starts the automated testing process. Afterwards, as can be seen in figure \ref{fig:mainscreen}, the main page displays the progress of the build and test steps as well as an overview of test results once the testing process is finished. As can be seen in the screenshot, the testing process consists of the following five steps:

\begin{enumerate}
    \item \textit{Select Files}: The user can optionally select which files of the project to include or exclude for testing. By default, all source code files (excluding header files) will be tested.
    \item \textit{Prebuild}: The clang prebuild step is run for each selected file, performing standard preprocessing and creating a single file containing the original source code as well as all code included from other files.
    \item \textit{Select Functions}: The user can optionally select individual functions to be included or excluded for testing. By default, all functions will be tested.
    \item \textit{Build}: For each test unit (i.e. each file), Coyote \CC\ automatically creates a test harness including drivers and stub functions, and, where needed to improve testability, some transformations are applied on both source and LLVM IR level without changing program semantics. The LLVM IR code is then compiled to an executable test binary.
    \item \textit{Test}: Test cases are iteratively generated and executed by the concolic execution engine.
\end{enumerate}

Once again, it should be noted that the entire build and test process is usually executed fully automatically and user intervention is only necessary when specific project files or individual functions should be excluded from testing.

\subsection{Inspecting Test Results}

\begin{figure}
    \centering
    \includegraphics[width=\linewidth]{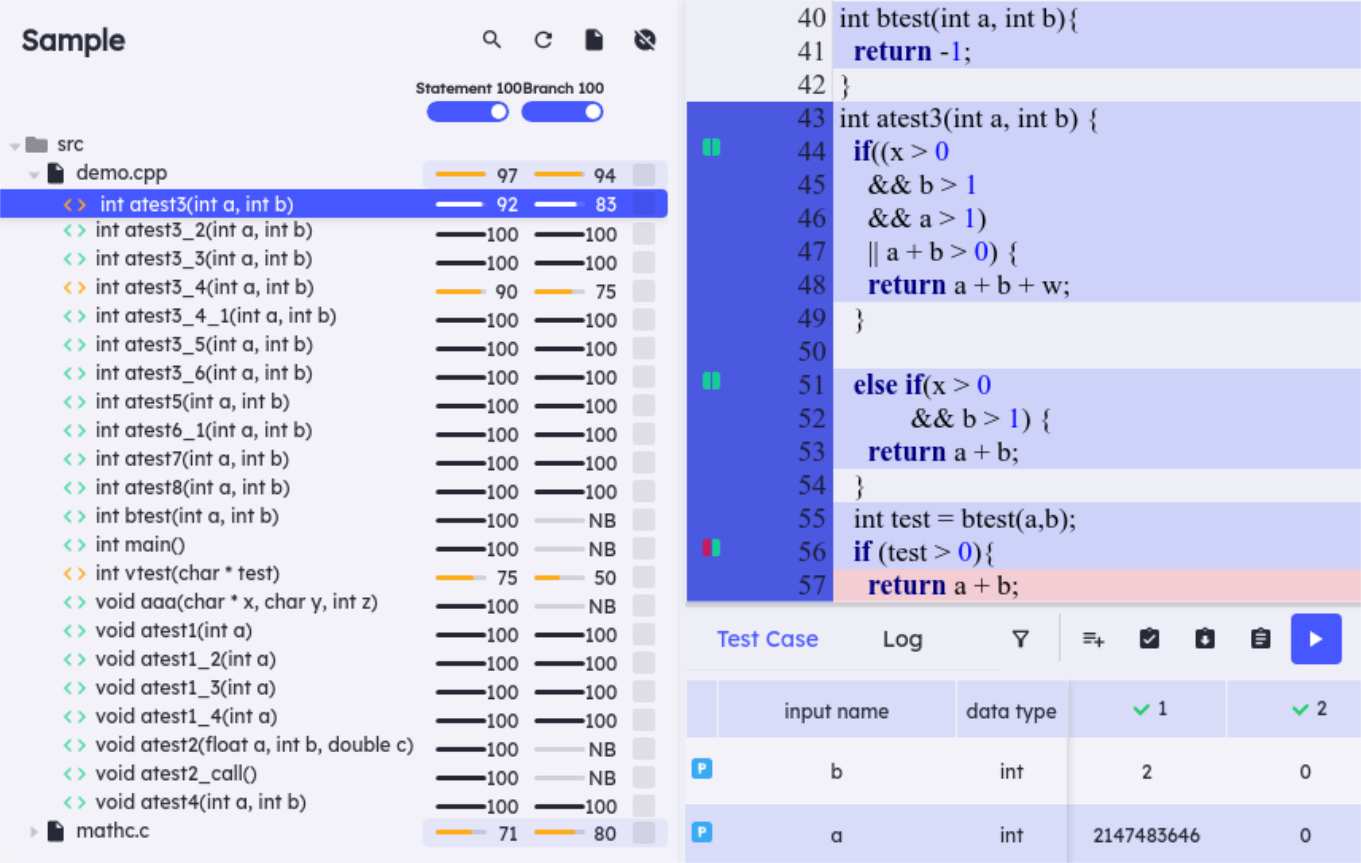}
    \caption{Results page with test cases and coverage info}
    \label{fig:resultsscreen}
\end{figure}

\begin{table*}[!ht]
\caption{Coverage and test time results for CITRUS and Coyote C++}
\begin{center}
\begin{tabular}{crrr|ccc|ccrc}
\hline
\multicolumn{4}{c|}{Project} & \multicolumn{3}{c|}{CITRUS} & \multicolumn{4}{c}{Coyote \CC}  \\ \hline
\multicolumn{1}{c|}{Name} & \multicolumn{1}{c|}{LOC} & \multicolumn{1}{c|}{\#Stmt} & \multicolumn{1}{c|}{\#Branch} & \multicolumn{1}{c|}{Stmt} & \multicolumn{1}{c|}{Branch} & \begin{tabular}[c]{@{}c@{}}Time\\ (hh:mm:ss)\end{tabular} & \multicolumn{1}{c|}{Stmt} & \multicolumn{1}{c|}{Branch}  & \multicolumn{1}{c|}{\begin{tabular}[c]{@{}c@{}}Test \\ cases\end{tabular}} & \begin{tabular}[c]{@{}c@{}}Time\\ (hh:mm:ss)\end{tabular} \\\hline \hline
\multicolumn{1}{c|}{JsonBox} & \multicolumn{1}{r|}{1,490} & \multicolumn{1}{r|}{778} & 342 & \multicolumn{1}{c|}{94.20\%} & \multicolumn{1}{c|}{79.10\%} & 16:12:00  & \multicolumn{1}{c|}{99.87\%} & \multicolumn{1}{c|}{98.83\%}  & \multicolumn{1}{r|}{1,030} & 00:01:42 \\ \hline
\multicolumn{1}{c|}{hjson-cpp} & \multicolumn{1}{r|}{4,497} & \multicolumn{1}{r|}{1,622} & 762 & \multicolumn{1}{c|}{80.20\%} & \multicolumn{1}{c|}{70.20\%} & 20:30:00 & \multicolumn{1}{c|}{92.29\%} & \multicolumn{1}{c|}{90.94\%} & \multicolumn{1}{r|}{7,075} & 00:04:44 \\ \hline
\multicolumn{1}{c|}{JSON Voorhees} & \multicolumn{1}{r|}{12,486} & \multicolumn{1}{r|}{2,281} & 688 & \multicolumn{1}{c|}{76.70\%} & \multicolumn{1}{c|}{50.30\%} & 20.48:00 & \multicolumn{1}{c|}{92.42\%} & \multicolumn{1}{c|}{87.79\%} & \multicolumn{1}{r|}{2,048} & 00:05:38 \\ \hline
\multicolumn{1}{c|}{JsonCpp} & \multicolumn{1}{r|}{9,691} & \multicolumn{1}{r|}{2,787} & 1,148 & \multicolumn{1}{c|}{95.40\%} & \multicolumn{1}{c|}{60.70\%} & 21.42:00 & \multicolumn{1}{c|}{92.25\%} & \multicolumn{1}{c|}{88.76\%} & \multicolumn{1}{r|}{4,250} & 00:19:21 \\ \hline
\multicolumn{1}{c|}{jvar} & \multicolumn{1}{r|}{5,373} & \multicolumn{1}{r|}{1,316} & 608 & \multicolumn{1}{c|}{84.50\%} & \multicolumn{1}{c|}{69.70\%} & 19:18:00 & \multicolumn{1}{c|}{96.05\%} & \multicolumn{1}{c|}{94.08\%} & \multicolumn{1}{r|}{1,068} & 00:05:35 \\ \hline
\multicolumn{1}{c|}{RE2} & \multicolumn{1}{r|}{27,687} & \multicolumn{1}{r|}{7,053} & 3,485 & \multicolumn{1}{c|}{80.20\%} & \multicolumn{1}{c|}{62.40\%} & 22:42:00 & \multicolumn{1}{c|}{89.08\%} & \multicolumn{1}{c|}{84.16\%} & \multicolumn{1}{r|}{7,574} & 00:10:35 \\ \hline
\multicolumn{1}{c|}{TinyXML-2} & \multicolumn{1}{r|}{5,627} & \multicolumn{1}{r|}{1,383} & 483 & \multicolumn{1}{c|}{59.50\%} & \multicolumn{1}{c|}{49.10\%} & 16.48:00  & \multicolumn{1}{c|}{90.96\%} & \multicolumn{1}{c|}{89.03\%} & \multicolumn{1}{r|}{1,607} & 00:03:35 \\ \hline
\multicolumn{1}{c|}{yaml-cpp} & \multicolumn{1}{r|}{78,290} & \multicolumn{1}{r|}{3,038} & 1,299 & \multicolumn{1}{c|}{80.60\%} & \multicolumn{1}{c|}{63.00\%} & 17:12:00 & \multicolumn{1}{c|}{97.07\%} & \multicolumn{1}{c|}{95.69\%} & \multicolumn{1}{r|}{3,340} & 00:05:14 \\ \hline \hline
\multicolumn{1}{c|}{Total} & \multicolumn{1}{r|}{145,141} & \multicolumn{1}{r|}{20,258} & 8,815 & \multicolumn{1}{c|}{81.41\%} & \multicolumn{1}{c|}{63.06\%} & 155:12:00 & \multicolumn{1}{c|}{92.34\%} & \multicolumn{1}{c|}{88.85\%} & \multicolumn{1}{r|}{27,992} & 00:56:24 \\ \hline
\end{tabular}
\label{tab:cov}
\end{center}
\end{table*}

Once the testing process for a project has finished, the results page of Coyote \CC\ shows detailed information about the generated test cases and achieved coverage results. As can be seen in figure \ref{fig:resultsscreen}, the left side of the page displays the statement and branch coverage results for each tested function individually as well as combined for each tested file. Selecting a function or file in this list shows its code on the right, where highlighting in and next to the code indicates precisely which lines and branches were covered by the generated test cases. These test cases can in turn be inspected in the bottom right part of the page, which shows the concrete values that were assigned to various symbols (i.e. program inputs) for each test case.

To deal with cases were Coyote \CC\ is unable reach satisfactory coverage results through automated testing, our tool also supports the manual addition of test cases as well as manual modifications of the code for driver and stub functions. We do not go into more detail about these features here as this paper focuses on the automated testing capabilities of Coyote \CC, but more information about manual improvement of test results can be found in the user manual provided alongside the Coyote \CC\ demo.

\section{Evaluation}

In order to demonstrate the improvement of testing performance achieved by Coyote \CC\ over the current state of the art of automated unit testing for \CC, we perform an evaluation where we compare the testing performance in terms of achieved coverage\footnote{As mentioned previously, the achieved code coverage is the main indicator for the quality of automated test case generation.} and time consumption between Coyote \CC\ and the CITRUS~\cite{Citrus} tool. We would have liked to include more tools into the comparison, but unfortunately other existing tools were either not publicly available, or, in the case of UTBot~\cite{Utbot}, were found to be unable to test any non-trivial code. Nonetheless, CITRUS is a rather recent tool reporting quite decent coverage results, so we still consider this comparison to be a meaningful evaluation. As CITRUS is no longer publicly available, we settled for using its previously published performance data as the baseline for our comparison by executing Coyote \CC\ on the same test set. In the course of this evaluation, Coyote \CC\ was executed on a Ubuntu 20.04 system equipped with a 24 core Intel i7-13700 and 64GB of RAM.

Table \ref{tab:cov} shows information about the test projects and the achieved testing performance of Coyote \CC\ and CITRUS, both in terms of yielded statement and branch coverage as well as time consumption. The statement counts are generally a lot lower than the physical lines of code because we only consider executable statements and header files as well as files only containing test code were excluded from automated testing. It should also be noted at this point that CITRUS only considers public functions as test targets, whereas Coyote \CC\ conducts automated testing for all functions regardless of access specifiers. Furthermore, while CITRUS originally reported multiple coverage results for different configurations, we only included the best results for each project respectively.

For CITRUS, the testing process took a total of about 155 hours to execute and the reported test coverage totals up to 81.41\% statement and 63.06\% branch coverage. Coyote \CC\ on the other hand was able to yield 92.34\% statement and 88.85\% branch coverage in only around 56 minutes. Therefore, Coyote \CC\ produces 10.93\% and 25.79\% higher statement and branch coverage with a test time that is two orders of magnitude faster than CITRUS. Despite the different test environments used for the two tools, this large of a difference clearly indicates that Coyote \CC\ produces significantly better coverage results in drastically less time than CITRUS. Additionally, the coverage results produced by Coyote \CC\ as well as the achieved testing speed of over 20,000 statements per hour once again fulfill the criteria for practicality of automated unit testing that we recently proposed: yielding at least 90\% statement coverage and 80\% branch coverage with a testing speed of more than 10,000 statements per hour.

\section{Conclusion}

We have presented Coyote \CC, a fully automated unit testing tool that features one-click automation of test case generation achieving coverage high enough to be practical for industrial use. Coyote \CC\ handles the well-known complexity of \CC\ syntax by generating sophisticated test drivers and function stubs, and then employs LLVM-based concolic execution to produce input data for concrete test cases. 
Coyote \CC\ also allows users to add supplementary test inputs and write specialized custom drivers and stubs for cases where automated testing is unable to reach 100\% coverage. In order to minimize the need for such user involvement, we plan to devise advanced methods to generate specially tailored drivers and stubs for functions with intricately structured inputs, such as using static analysis to infer the range of accessed indexes for arrays with unspecified size or to find the minimum number of loop iterations necessary to cover all statements and branches in a function.

\bibliographystyle{IEEEtran}
\bibliography{bibliography}

\begin{thebibliography}{10}
\providecommand{\url}[1]{#1}
\csname url@samestyle\endcsname
\providecommand{\newblock}{\relax}
\providecommand{\bibinfo}[2]{#2}
\providecommand{\BIBentrySTDinterwordspacing}{\spaceskip=0pt\relax}
\providecommand{\BIBentryALTinterwordstretchfactor}{4}
\providecommand{\BIBentryALTinterwordspacing}{\spaceskip=\fontdimen2\font plus
\BIBentryALTinterwordstretchfactor\fontdimen3\font minus \fontdimen4\font\relax}
\providecommand{\BIBforeignlanguage}[2]{{%
\expandafter\ifx\csname l@#1\endcsname\relax
\typeout{** WARNING: IEEEtran.bst: No hyphenation pattern has been}%
\typeout{** loaded for the language `#1'. Using the pattern for}%
\typeout{** the default language instead.}%
\else
\language=\csname l@#1\endcsname
\fi
#2}}
\providecommand{\BIBdecl}{\relax}
\BIBdecl

\bibitem{klee}
C.~Cadar, D.~Dunbar, D.~R. Engler \emph{et~al.}, ``{KLEE}: Unassisted and automatic generation of high-coverage tests for complex systems programs.'' in \emph{OSDI}, vol.~8, 2008, pp. 209--224.

\bibitem{cute}
K.~Sen, D.~Marinov, and G.~Agha, ``{CUTE}: A concolic unit testing engine for {C},'' \emph{ACM SIGSOFT Software Engineering Notes}, vol.~30, no.~5, pp. 263--272, 2005.

\bibitem{EvoSuite}
\BIBentryALTinterwordspacing
G.~Fraser and A.~Arcuri, ``A large-scale evaluation of automated unit test generation using {E}vo{S}uite,'' \emph{ACM Trans. Softw. Eng. Methodol.}, vol.~24, no.~2, dec 2014. [Online]. Available: \url{https://doi.org/10.1145/2685612}
\BIBentrySTDinterwordspacing

\bibitem{Transformer}
\BIBentryALTinterwordspacing
M.~Tufano, D.~Drain, A.~Svyatkovskiy, S.~K. Deng, and N.~Sundaresan, ``Unit test case generation with transformers,'' \emph{ArXiv}, vol. abs/2009.05617, 2020. [Online]. Available: \url{https://api.semanticscholar.org/CorpusID:221655653}
\BIBentrySTDinterwordspacing

\bibitem{schafer2023empirical}
M.~Sch{\"a}fer, S.~Nadi, A.~Eghbali, and F.~Tip, ``An empirical evaluation of using large language models for automated unit test generation,'' \emph{arXiv preprint arXiv:2302.06527}, 2023.

\bibitem{jcute}
K.~Sen and G.~Agha, ``Cute and jcute: Concolic unit testing and explicit path model-checking tools,'' in \emph{Computer Aided Verification}, T.~Ball and R.~B. Jones, Eds.\hskip 1em plus 0.5em minus 0.4em\relax Berlin, Heidelberg: Springer Berlin Heidelberg, 2006, pp. 419--423.

\bibitem{Maist}
Y.~Kim, D.~Lee, J.~Baek, and M.~Kim, ``Concolic testing for high test coverage and reduced human effort in automotive industry,'' in \emph{2019 IEEE/ACM 41st International Conference on Software Engineering: Software Engineering in Practice (ICSE-SEIP)}.\hskip 1em plus 0.5em minus 0.4em\relax IEEE, 2019, pp. 151--160.

\bibitem{dart}
P.~Godefroid, N.~Klarlund, and K.~Sen, ``{DART}: Directed automated random testing,'' in \emph{Proceedings of the 2005 ACM SIGPLAN conference on Programming language design and implementation}, 2005, pp. 213--223.

\bibitem{Pex}
N.~Tillmann and J.~de~Halleux, ``Pex--white box test generation for .net,'' in \emph{Tests and Proofs}, B.~Beckert and R.~H{\"a}hnle, Eds.\hskip 1em plus 0.5em minus 0.4em\relax Berlin, Heidelberg: Springer Berlin Heidelberg, 2008, pp. 134--153.

\bibitem{cuter}
\BIBentryALTinterwordspacing
A.~Giantsios, N.~Papaspyrou, and K.~Sagonas, ``Concolic testing for functional languages,'' \emph{Science of Computer Programming}, vol. 147, pp. 109--134, 2017, selected and Extended papers from the International Symposium on Principles and Practice of Declarative Programming 2015. [Online]. Available: \url{https://www.sciencedirect.com/science/article/pii/S0167642317300837}
\BIBentrySTDinterwordspacing

\bibitem{jalangi}
\BIBentryALTinterwordspacing
K.~Sen, S.~Kalasapur, T.~Brutch, and S.~Gibbs, ``Jalangi: A selective record-replay and dynamic analysis framework for javascript,'' in \emph{Proceedings of the 2013 9th Joint Meeting on Foundations of Software Engineering}, ser. ESEC/FSE 2013.\hskip 1em plus 0.5em minus 0.4em\relax New York, NY, USA: Association for Computing Machinery, 2013, p. 488–498. [Online]. Available: \url{https://doi.org/10.1145/2491411.2491447}
\BIBentrySTDinterwordspacing

\bibitem{Citrus}
R.~S. Herlim, Y.~Kim, and M.~Kim, ``{CITRUS}: Automated unit testing tool for real-world {C++} programs,'' in \emph{2022 IEEE Conference on Software Testing, Verification and Validation (ICST)}, 2022, pp. 400--410.

\bibitem{Utbot}
D.~Ivanov, A.~Babushkin, S.~Grigoryev, P.~Iatchenii, V.~Kalugin, E.~Kichin, E.~Kulikov, A.~Misonizhnik, D.~Mordvinov, S.~Morozov \emph{et~al.}, ``{UnitTestBot}: Automated unit test generation for {C} code in integrated development environments,'' in \emph{2023 IEEE/ACM 45th International Conference on Software Engineering: Companion Proceedings (ICSE-Companion)}.\hskip 1em plus 0.5em minus 0.4em\relax IEEE, 2023, pp. 380--384.

\bibitem{Coyote}
S.~Rho, P.~Martens, S.~Shin, Y.~Kim, H.~Heo, and S.~Oh, ``Coyote c++: An industrial-strength fully automated unit testing tool,'' 2023.

\bibitem{SymExecSurvey}
\BIBentryALTinterwordspacing
R.~Baldoni, E.~Coppa, D.~C. D’elia, C.~Demetrescu, and I.~Finocchi, ``A survey of symbolic execution techniques,'' \emph{ACM Comput. Surv.}, vol.~51, no.~3, may 2018. [Online]. Available: \url{https://doi.org/10.1145/3182657}
\BIBentrySTDinterwordspacing

\end{thebibliography}

\end{document}